\def\year{2022}\relax
\documentclass[letterpaper]{article} 
\usepackage{aaai22}  
\usepackage{times}  
\usepackage{helvet}  
\usepackage{courier}  
\usepackage[hyphens]{url}  
\usepackage{graphicx} 
\usepackage{natbib}  
\usepackage{caption} 
\DeclareCaptionStyle{ruled}{labelfont=normalfont,labelsep=colon,strut=off} 
\usepackage{algorithm}
\usepackage{algorithmic}

\usepackage{newfloat}
\usepackage{listings}
\floatstyle{ruled}
\newfloat{listing}{tb}{lst}{}
\floatname{listing}{Listing}
\title{Local Justice and the Algorithmic Allocation of Scarce Societal Resources}
\author{
    Sanmay Das
    }
\affiliations{
    George Mason University\\

    sanmay@gmu.edu
}

\begin{document}

\maketitle

\begin{abstract}
  AI is increasingly used to aid decision-making about the allocation
  of scarce societal resources, for example housing for homeless
  people, organs for transplantation, and food donations. Recently,
  there have been several proposals for how to design objectives for
  these systems that attempt to achieve some combination of fairness,
  efficiency, incentive compatibility, and satisfactory aggregation of
  stakeholder preferences. This paper lays out possible roles and
  opportunities for AI in this domain, arguing for a closer engagement
  with the political philosophy literature on local justice, which
  provides a framework for thinking about how societies have over time
  framed objectives for such allocation problems. It also discusses
  how we may be able to integrate into this
  framework the opportunities and risks opened up by the ubiquity of
  data and the availability of algorithms that can use them to
  make accurate predictions about the future.
\end{abstract}

\section{Introduction}

In a recent paper, \citet{freedman2020adapting} remark that
``efficient and fair allocation of limited resources is a classical
problem in economics and computer science.'' Their specific
application is to kidney exchange, a central example of the kinds of
domains we are concerned with in this paper, but it also captures many
important issues in the development of how we think about the problem
today. Computer Science has been concerned for a long time with
algorithms that allow for efficient and fair allocation of limited
resources (learning about job allocation in time-sharing computers is
a long-time staple of operating systems courses, for
example). Economics is sometimes defined as precisely the study of
the allocation of scarce resources. As a result, most of our ideas for
AI-enabled allocation of scarce resources 
have been driven by the histories of these two fields. This has
significant consequences, because it means the measures we develop
algorithms to optimize are those that have a pride of place in
economics -- most of the time utilitarian (or additive) social
welfare, but in other cases Rawlsian (max-min) or Nash
(multiplicative) social welfare.

However, we have also been building real social and political
institutions to allocate scarce resources for centuries, and a central
insight is that they often prioritize in ways that do not correspond
directly with any of these measures of social welfare. For example,
triage in battlefield medicine or rationing of healthcare in
emergencies prioritizes greedily by predicted improvement, which can
have consequences different from either efficient allocation or
max-min allocation.
While economics may be the study of
the allocation of scarce resources, in the words of
Harold Lasswell, politics is ``who gets what, when, and how.''
Thus it is appropriate to turn to political science. The political
philosophy of allocation of scarce resources is studied under the
moniker of \emph{local justice} \cite{elster1992local}, which
systematically considers the question of how institutions allocate
scarce resources and necessary burdens (for example, through
lotteries, principles of greatest need, best outcome, or most ``value
added''). 

While there has been burgeoning interest in defining the objectives of
AI systems by taking multiple stakeholders into account
\cite{freedman2020adapting,lee2019webuildai}, the main theme of this paper is
that analyzing AI systems that endeavor to fairly and efficiently
allocate scarce resources through the lens of local justice can
greatly clarify both the objectives of system design and the parts of
the pipeline where algorithmic and data-driven techniques can be
particularly helpful. I start by introducing the setting and
describing four big questions that should be answered by any allocation
system. After providing background on related research in machine 
learning and fair division, I introduce a formalization of some of the
principles of local justice, and then discuss how the four questions
can be thought of in that framework, and the roles that AI can play in
helping to better assess and design these allocation systems.

\section{Setting}

For the purposes of this paper, the problems we are interested in involve the allocation of resources that are
(1) controlled or regulated by society; and (2) scarce; we focus on
settings where data-driven or algorithmic decision-making is
feasible. For various reasons, we have decided that market mechanisms
are inappropriate for these settings \cite{elster1992local,roth2007repugnance,currie2008transfers}. Examples include organs for transplantation,
resources for homeless populations, and spaces in elite public
schools, among others. Note that decisions about the appropriateness
of market mechanisms can vary across settings -- for example,
markets for kidneys exist in Iran. A few other notes: settings can be
dynamic (e.g. organ transplantation) or static (e.g. batch matchings
of students to schools), and it could be possible to assign different
individuals to different types of resources (for example, the level of
services provided by homelessness service providers can vary in
intensity).

I will argue, through the rest of this paper, that in order to design
systems that are improved by the use of AI (and more generally, to
consider the benefits and harms of using AI in these domains), we need
to think clearly and specifically about a set of related questions.
(1) How do we define and quantify desirable outcomes, in terms of
efficiency, equity, justice, or fairness? (2) How do we predict outcomes for
heterogeneous individuals and households under different feasible
allocations? (3) How do we optimize allocation of scarce resources
to achieve the best population-level outcomes under constraints
defined by our notions of justice or fairness?
(4) What can we say about the incentives created by the overall
system, and the potential for manipulation or negative long-term
outcomes of deployment, considering the preferences of participants?

By specifically engaging elements of each of these questions, we can better consider the ramifications of our technologies --
how, precisely, do they help or hurt compared with current practice?
-- rather than being seen as technological solutionists by the
stakeholders we must engage.

\section{Background}

It is certainly not a novel observation that we need to think
carefully about what it is that we are trying to optimize. Just to
give a couple of recent examples, \citet{conitzer2019designing} discusses the
importance of appropriately designing preferences and optimization
goals for AI agents, while a core argument of \citet{o2020near} is that many of the problems of ``near-term AI'' (defined as
expert systems that replace human decision-makers) are driven by a
mismatch between the performance metrics of the AI (constructed by the
algorithm designers) and the true objectives of
stakeholders. Nevertheless, it is useful to get a sense of where the
academic community has gone in response to these concerns.

\subsection{Fairness in machine learning}

A common trope is that the first objective of the engineer is simply
to optimize a given objective function. Indeed, standard metrics, for
example, accuracy, area under the ROC curve, or return on investment,
still drive much research, so the first instinct in many applications
of machine learning in society has been to define societal problems in
a manner amenable to analysis through the lens of metrics like these.
In the recent past, we have learned how machine learning systems that
are ``in the loop'' of human decision-making can have significant
unintended consequences. Examples abound: in some cases, instead of
reducing crime rates, predictive policing results in more false
arrests as police misinterpret algorithmic predictions of suspects as
evidence \cite{saunders2016predictions}. In a number of situations,
data-driven allocations have unintentionally introduced systematic
biases that perpetuate inequities, such as racial disparities in
credit lending, hotspot policing, and crime sentencing
\cite{ensign2017runaway,pleiss2017fairness,corbett2017algorithmic}.

The response of the ML community, while not entirely uniform, has
largely revolved around a call for algorithms to satisfy various
fairness metrics
\cite{dwork2012fairness,kusner2017counterfactual,hardt2016equality}. However,
there has been pushback against this from various
perspectives. Notably there have been some impossibility results,
showing that several different fairness criteria that all seem
intuitively reasonable cannot be satisfied simultaneously
\cite{kleinberg2016inherent,pleiss2017fairness,feller2016computer}. \citet{corbett2018measure} discuss the statistical limitations of various fairness
criteria, and argue that formal fairness criteria may ``harm the very
groups they are meant to protect'' and advocate instead for treating
similarly risky individuals similarly, based on the best risk metrics
available. \citet{green2018myth} say that the
methodological reliance of machine learning on standard techniques,
metrics, and datasets makes it ill-suited to address political and
ethical considerations in the deployment of algorithms in socially
important contexts. They go on to call for the
process of democratic deliberation as much as technical analysis in
such deployments.

There is also a rich literature on \emph{fair division} \cite[e.g.]{bouveret2008efficiency,chen2013truth}
that is related, but the connections are, somewhat surprisingly, only
beginning to be explored. In typical fair division problems, the
analysis is from the perspective of treating all agents as equal
priority and focusing on the efficiency and fairness guarantees that
can be made with respect to agent preferences. In the types of domains
we are looking at, the questions of who receives priority and why are
much more central. 


\section{Social policy and local justice}
In the domain of allocation of scarce resources to individuals (and
households, etc.) in need, we do not have to reinvent the wheel
in order to examine the effects of different ways of setting social
optimization goals. Institutions such as organ donation policy-making
bodies and draft boards (among many others) have long grappled with
the question of how best to allocate scarce resources (or necessary
burdens). Political philosophers have discerned from these cases a set
of useful underlying principles of what they call local justice. Our
discussion here largely follows that of \citet{elster1992local}.

\paragraph{Local vs.\ global justice}
To make one distinction clear, local justice is
distinguished from theories of global justice and individual rights
(notably utilitarianism, Rawl's theory, and Nozick's libertarian
theory) by the local nature of decision-making. The ethics of deciding
how to allocate scarce resources in one setting does not carry over to
others necessarily, and a series of decisions deemed to be locally
just may lead to global problems for a particular group. Our concern
here is with considerations of ethical decision-making by local
institutions. 

\subsection{Principles of local justice}

Elster categorizes the principles of local justice in several ways. Let us briefly discuss principles for
allocation of scarce resources that have been used in society that
will not be our focus. These include allocation based on status (quota
systems in general), age and gender (women and children first on the
lifeboats), waiting time (queueing systems), power and influence
(legacy admissions), or lotteries.

If we restrict ourselves to settings in which AI or data-driven
decision-making can have most impact, we are most interested in
principles that take into account specific properties of individuals,
and also the interaction between the individual and the allocated
resource. Three types of local decision-making based on the welfare of
individuals are prevalent in institutions that allocate scarce
resources. In order to present these, first let us assume that
individual welfare levels can be reduced to a single-dimension, call
it $w$.

\noindent (1) Minimum pre-allocation $w$: This is the principle of allocating
  to those with the greatest need. The homelessness system in most of the United States is a good example of
  a system that mostly works on this principle, with explicit
  determination and prioritization of the most vulnerable, sometimes
  based on a score, and sometimes on individual discretion of the case
  officer. Another example is in cadaveric organ transplantation,
  where the sickest patients get highest priority, often based on a
  complex scoring mechanism.
  
\noindent (2) Maximum post-allocation $w$: This principle allocates
  resources to those who will be best off after allocation. For
  example, some elite public magnet schools may select those
  who are already extremely gifted and would have the
  highest post-schooling quality (even if the school does not
  end up contributing much to that final quality itself).

  \noindent (3) Greatest increase in $w$: This is the principle of
  allocating to those who would get the greatest ``value added'' from
  the resource, measured by the difference in post- and pre-allocation
  levels of $w$ (note that this difference could be stochastic rather
  than deterministic). An example is in emergency medical triage in
  wars, or when there are insufficient ICU beds or ventilators
  available in hospitals and care must be rationed. In these settings,
  those who receive treatment and attention are typically those in the
  middle-range (leaving aside those who will recover well without
  attention and those who are too critical to be saved).

It is certainly not the case that one of these is better than the
others as a criterion in all situations. Indeed it is worth thinking
about societal objectives again, with the example of patients on wait
lists for organs from deceased donors. It is common practice to
prioritize sicker patients (corresponding to the first principle
above) even though overall outcomes may be better if one were to
transplant less sick patients earlier in their time on the wait
list. However, this creates the societal feeling that one could be
abandoned, which is considered harmful to social
well-being and cohesion. 

\section{Roles for AI}

Using the framework of local justice, and in particular of allocation
systems based on properties of individuals, we can articulate possible
roles for AI by going back to the four defining questions for an
allocation system posed above.

\subsection{Defining and quantifying desirable outcomes}

While most of the literature examines the ``traditional'' utilitarian
or Rawlsian notions of welfare, there has been increasing interest in
considering different possibilities. The ``moral machines'' project
was among the first to crowdsource moral judgments about algorithmic
decisions in the context of ethical dilemmas that could be faced by
autonomous vehicles \cite{awad2020crowdsourcing}. There has been
recent work on turning such judgments, from experts or from the
general public, into objective functions. Notably, \citet{lee2019webuildai}
describe a system that allows stakeholders to construct computational
models representing their views, and the models then vote in order to
create what the authors call ``algorithmic policy.'' One of the
benefits, compared to objective functions designed by the algorithm
designers, is that this engages stakeholders and increases
buy-in. \citet{freedman2020adapting} propose a principled methodology
for eliciting stakeholder preferences for which attributes of
individuals should be considered for prioritization of kidney
transplant recipients, and propose a method for estimating weights
using these preferences. Interestingly, in both of these cases, there
is a combination of machine learning for preference learning, and some
form of social choice for preference aggregation. These types of
ideas highlight the promise of AI to go beyond what we have been able
to do societally and come up with methods that could lead to
greater social acceptance and more participatory decision-making in
these spheres.

Another area where AI has been valuable is in understanding the
overall effects of optimizing different welfare criteria through ideas
like the price of fairness, which examines the efficiency loss due to
implementation of a fairness criterion in allocation \cite{bertsimas2011price,caragiannis2012efficiency}.  For example, \citet{nguyen2021scarce} consider a
prioritization method that implements the triage principle
of greatest increase in $w$. They study the price of fairness and show
that its price of fairness is often lower than that of the
most-vulnerable-first principle. They argue that this could partly
explain situations in which society has agreed that the triage
principle should be used, e.g. short-term emergency medical situations
like those that could occur in a pandemic, in spite of the problem of
abandonment.  In this type of work, computing acts, in the words of \citet{abebe2020roles} ``as a formalizer, [shaping] how social problems are explicitly defined --- changing how those problems, and possible responses to them, are understood.''

There are many risks and caveats to remain aware of. While some may be
unanticipated, a few can be seen from the start, and we discuss two
here. First: some of the major problems in bias of machine learning
systems have arisen as a result of them replicating human biases that
we do not want to sustain. Does learning preferences from stakeholders
and humans potentially perpetuate injustices? This can vary between
contexts, but the famous case of the Seattle ``God committee''
which decided who would have access to dialysis treatments in the
1960s is a classic cautionary tale. As \citet{levine2009seattle} says ``the
committee relied heavily on criteria of social worth heavily weighted
toward economic status, reflecting their own values and biases.''
One of the potential \emph{benefits} of AI
is in the potential objectivity of its goals, but nevertheless, this must
be in consultation with stakeholders (the veneer of objectivity can
also encode biases, of course).

Second, we need to be careful in translating between social objectives
and mathematical formalizations. We often benchmark to efficient, or
utility maximizing allocations, whether in terms of the price of
fairness or not. However, even in some social allocation problems
where the goal at first appears to be utilitarian efficiency, it may not be. For
example, prioritization by maximum post-allocation welfare is not the
same thing as the utilitarian assignment in a system with multiple
interventions.

\subsection{Predicting outcomes under different allocations}

A second part of the local justice framework in which AI, and in
particular, machine learning (ML), has potential is in measuring and
predicting the welfare level $w$. While an
impossible task in general, in many real-world cases we do use
proxies for $w$. For example, in medical interventions we may use
Quality Adjusted Life Years (QALY) as a
measure, in homelessness services the VI-SPDAT score (note concerns
about validity) has been used to measure vulnerability, and in liver
transplantation the MELD score is used to measure severity of liver
disease. In the latter two cases, the score is considered inversely
related to welfare.

One interesting use of ML is to better capture
what such scores seek to measure. In homelessness services, for
example, the vulnerability score is often intended to measure how
well an individual or household would do without receiving support. In
child welfare, one may want to measure any number of negative outcomes
(inability to make progress in school, interactions with
the criminal justice system or need for emergency medical care). By using ML methods on administrative data,
one could predict the probability of successfully remaining housed, or
graduating school on time, and these measures
may be more directly related to the outcomes we care about
than the scores we currently construct. 

The prediction problem is in some ways even more interesting, because
it asks the question of how well individual $A$ (defined by some
characteristics that we can think of as a feature vector) would do
when given interventions $\alpha, \beta, \gamma, \ldots$. This is more
complex because it asks about the interaction of an individual (or
household) with an intervention. While there is much to be done in
this area, recently different approaches to this have been taken. In
the domain of matching refugees to cities in which they are likely to
do well (so $A$ is a refugee family and the Greek letters are
different cities), \citet{bansak2018improving} take advantage of prior random allocation
to learn good models, while \citet{kube_allocating_2019} estimate
counterfactual probabilities using BART in the case of allocation of
homelessness services. 

Many of the usual concerns with ML surface when it is
used in these domains. Are we learning the right model, and making
predictions in ways that are fair to different subgroups, for example
\cite{kearns2018preventing}? This could be compounded in two separate ways. First,
there are many domains where human behavior is fundamentally not very
predictable compared with other ML tasks, with low AUC
values the norm. Second, the counterfactual problem is particularly difficult. There may be a paucity of data on
individuals or households of certain types receiving certain services
in the past (whether because of biases or other reasons), making
models less valid for those populations.

\subsection{Optimization with fairness constraints}

The third question raises interesting algorithmic challenges. Given a plausible, but perhaps ``soft'' notion of fairness that
does not yield a strict priority ordering that must be respected, and
also a social goal, how can we optimize towards both objectives? This
is a fascinating, and methodologically challenging, problem with
a strong practical basis. For example, those fleeing domestic violence
are prioritized for spaces in homeless shelters and the
immunosuppressed for vaccines. There is exciting work in this
space. For example \citet{mcelfresh2018balancing} design a rule for balancing a utilitarian
objective with preferences for one group over another in kidney exchange, while \citet{azizi2018designing} consider a
formulation that optimizes efficiency while satisfying fairness
constraints specified by a policy-maker in the context of providing
resources to homeless youth. There is plenty to be done in this area,
where exact solutions that can be computed efficiently are typically hard to find.

\subsection{Preferences and incentives}

Thus far we have elided the question
of preferences. The implicit assumption is that the decision-maker
knows, or can acquire, sufficient information about each individual in
order to make decisions, and that individuals do not really have the
power to affect allocations, and would prefer some allocation to none
at all. However, we should also consider the incentives created by
such systems, as suggested by \citet{roth1993local} in a review of Elster's book
\emph{Local Justice}. Recent work in algorithmic game theory
and matching has begun to do so. Two examples are
illustrative. \citet{estornell2021incentivizing} look at a problem
motivated by people lying about self-reported attributes in order to
receive a more favorable score (as has been reported, for example, in
the context of homelessness services in LA). They show how one
can use audits of self-reported features after-the-fact in order to
incentivize truthful behavior, and also demonstrate that the scarce
resource setting poses fundamentally different challenges than the
setting where resources are not scarce but the institution would
benefit from setting some score threshold. \citet{aziz2021efficient}
consider the problem of allocating scarce healthcare resources when
individuals may have different eligibilities for the resources. For
example, individuals may become eligible for Covid-19 vaccination
under different eligibility categories. They design a mechanism for
efficient allocation that complies with eligibility requirements,
respects priorities within eligibility categories, and incentivizes
truthfulness in the sense that individuals never underreport
eligibility categories.

These examples are on the spectrum of the many interesting and
challenging problems that arise on the mechanism design side when
thinking about how to design end-to-end systems for allocation of
scarce societal resources. In addition to the other categories above,
this provides a set of possible ideas, and a personal sense of
priorities, for future research in the use of AI for such allocation
problems.

\section{Acknowledgments}

I am grateful to Frank Lovett for pointing me to Jon Elster's work, and to Veena Das, Jennifer Dlugosz, Patrick Fowler and Eugene Vorobeychik for many conversations that have helped frame my thinking on these topics. I also thank the AAAI Blue Sky track reviewers for various helpful suggestions. All errors and opinions are, of course, my own. This work has been supported in part by NSF awards 2127752, 2127754, and 1939677 and by Amazon through an NSF-Amazon FAI award.

\bibliography{bib11,sd-pubs-biber,homeless}

\end{document}